\documentclass[aps]{revtex4-2}
\usepackage{eurosym}
\usepackage{amssymb}
\usepackage{amsfonts}
\usepackage{amsmath}
\usepackage{graphicx}
\usepackage{bm}
\usepackage{braket}
\usepackage{mathtools}
\usepackage{textcomp}

\begin{document}

\title{Spontaneous symmetry breaking in continuous waves, dark solitons, and
vortices in linearly coupled bimodal systems}
\author{Hidetsugu Sakaguchi$^{1}$ and Boris A. Malomed$^{2,3}$}
\address{$^{1}$Department of Applied Science for Electronics and Materials, Interdisciplinary Graduate School of
Engineering Sciences, Kyushu University, Kasuga, Fukuoka 816-8580, Japan}
\address{$^{2}$Department of Physical Electronics, School of Electrical Engineering,
Faculty of Engineering, and Center for Light-Matter Interaction, Tel Aviv
University, P.O. Box 39040 Tel Aviv, Israel}
\address{$^{3}$Instituto de Alta Investigaci\'{o}n, Universidad de Tarapac\'{a}, Casilla 7D,
Arica, Chile}

\begin{abstract}
We introduce a model governing the copropagation of two components which
represent circular polarizations of light in the optical fiber with relative
strength $g=2$ of the nonlinear repulsion between the components, and linear
coupling between them. A more general system of coupled Gross-Pitaevskii
(GP) equations, with $g\neq 2$ and the linear mixing between the components,
is considered too. The latter system is introduced in its one- and
two-dimensional (1D and 2D) forms. A new finding is the spontaneous symmetry
breaking (SSB) of bimodal CW (continuous-wave) states in the case of $g>1$
(in the absence of the linear coupling, it corresponds to the immiscibility
of the nonlinearly interacting components). The SSB is represented by an
exact asymmetric CW solution. An exact solution is also found, in the case
of $g=3$, for stable dark solitons (DSs) supported by the asymmetric CW
background. For $g\neq 3$, numerical solutions are produced for stable DSs
supported by the same background. Moreover, we identify a parameter domain
where the fully miscible (symmetric) CW background maintains stable DSs with
the \textit{inner SSB} (separation between the components) in its core. In
2D, the GP system produces stable vortex states with a shift between the
components and broken isotropy. The vortices include ones with the
inter-component shift imposed by the asymmetric CW background, and states
supported by the symmetric background, in which the intrinsic shift
(splitting)) is exhibited by vortical cores of the two components.
\end{abstract}

\maketitle

\section{Introduction and the model}

Spontaneous symmetry breaking (SSB) is a ubiquitous feature of phenomenology
which occurs in a great variety of nonlinear systems. Various manifestations
of SSB have been studied in detail, theoretically and experimentally, in
nonlinear optics and photonics and in atomic BECs (Bose-Einstein
condensates) \cite{book}. A basic setting for the realization of SSB is
provided by linearly coupled dual-core (two-component) systems with
intrinsic (intra-core) nonlinearity, such as twin-core optical fibers
(a.k.a. nonlinear couplers) \cite{Jensen,Maier}, tunnel-coupled traps for
matter waves \cite{Oberthaler,Markus,HS}, microwave resonators \cite{Buks},
and some other physical systems. In these systems, SSB is initiated by
instability of symmetric two-component states against small perturbations
which break their symmetry. In their simplest form, the corresponding SSB
effects occur in the continuous-wave (CW) settings \cite{Snyder}. However,
CW states are subject to the modulational instability (MI) in the case of
the self-focusing intra-core nonlinearity, which tends to split them into
chains of bright solitons \cite{Agrawal}. Thus, much interest has been drawn
to the realization of SSB in two-component solitons carried by dual-core
optical fibers \cite{Wabnitz}-\cite{we}, see also review \cite{Peng-book}.
The conservative dual-core systems are modeled by systems of two
linearly-coupled nonlinear Schr\"{o}dinger (NLS) equations, while their
dissipative counterparts are represented by systems of coupled complex
Ginzburg-Landau equations, which predict SSB in two-component dissipative
solitons \cite{Sigler,new}.

The natural sign of the intrinsic nonlinearity in BEC is self-repulsive \cite%
{Pit}, which is possible too in defocusing nonlinear optical media \cite%
{Swartz,Kivsh}. Intrinsic instabilities take place in two-component
self-defocusing system if coefficients $g_{11,22}$ and $g_{12}\equiv g_{21}$
of the cross- and self-repulsion of the components satisfy the basic
immiscibility condition \cite{Mineev}, $g_{12}>\sqrt{g_{11}g_{22}}$. Below,
we consider the system with $g_{11}=g_{22}$, and define the relative
strength of the cross-repulsion as $g\equiv g_{12}/g_{11,22}$, in terms of
which the immiscibility condition amounts to $g>1$. In this case, spatially
or temporally periodic perturbations give rise to a specific form of MI,
which breaks the uniform CW state into a chain domain walls (DWs) separating
domains populated by the different immiscible components \cite{Agrawal-MI}.

The SSB phenomenology in linearly-coupled dual-core systems with intrinsic
defocusing nonlinearities was predicted in two-component gap solitons, that
are supported by such systems if both cores carry a spatially-periodic
potential \cite{Markus}, which is necessary to create gaps solitons as a
result of the interplay of this potential with the self-repulsion \cite{Peli}%
. The SSB between linearly coupled components in 1D and 2D systems combing
the self-defocusing and a trapping harmonic-oscillator potential was
addressed in Ref. \cite{we}, which included the consideration of 2D vortex
states with the broken symmetry. However, a possibility of SSB in
two-component systems combining the linear coupling between the components
with the self- and cross-repulsion in the free space has not been addressed
previously. The objective of the present work is to introduce such a
physically relevant system, produce SSB\ states in it, including CWs,
two-component one-dimensional (1D) dark solitons (DSs), and two-component 2D
vortices, in analytical and numerical forms, and investigate the stability
of these states.

The system is introduced in Section 2, in the form of the system of a
linearly-coupled Gross-Pitaevskii (GP) equations for wave functions $\phi
(x) $ and $\psi (x)$. General CW analytical solutions with the unbroken and
broken symmetry, as well as specific exact DS solutions, are found in
Section 3. The stable asymmetric CW\ states exist precisely in the case when
the immiscibility condition $g>1$ holds. Further, in the special case of $%
g=3 $, which implies strong immiscibility,we find an exact DS solution,
which connects two asymmetric CW states with opposite overall signs at $%
x=\pm \infty $. While this DS is maintained by the asymmetric CW background,
it keeps its intrinsic inversion symmetry, expressed by relation%
\begin{equation}
\phi (-x)=-\psi (x).  \label{phipsi}
\end{equation}%
For $g\neq 3$, we find more general exact DS asymmetric solutions pinned to
the attractive or repulsive P\"{o}schl-Teller (PT) potential, of the $%
\mathrm{sech}^{2}$ type \cite{PT} (although, unlike the exact DS solutions
found at $g=3$, they are unstable). Section 3 also reports the MI analysis
for the CW symmetric states at the SSB threshold, which demonstrates that
they are stable, hence the asymmetric CW states, produced by the SSB
bifurcation, inherit the modulational stability from the symmetric ones.

Systematically produced numerical results are reported in Section 4, for
both the symmetry-breaking DSs and their 2D counterparts, which, as it may
be naturally expected \cite{Swartz}, are two-component vortices. The 1D
numerical solutions are produced for a pair of DSs which are placed at
opposite positions in a domain of a large size, subject to periodic boundary
conditions, with the aim to preclude artifacts that may be induced by
boundary conditions. In particular, stable numerical solutions are produced
for the symmetry-breaking DSs in the general case, $g\neq 3$. A noteworthy
finding, revealed by the numerical solution and explained by a variational
approximation (VA), is that immiscibility\ (actually, a shift between the
components) may take place in the core of the 1D DSs or 2D vortices, even in
the case when the CW background which supports them remains fully miscible
(symmetric). We categorize this case as \textquotedblleft inner"
immiscibility, to distinguish it from the \textquotedblleft outer"
immiscibility, featured by the external CW background (which implies the
occurrence of immiscibility in the core region as well). The paper is
concluded by Section 5.

\section{The model}

In the scaled form, the 1D system of linearly coupled GP equations for
mean-wave functions $\phi $ and $\psi $ of the two components \ of the
binary BEC, with the repulsive self- and cross-interactions, is%
\begin{eqnarray}
i\phi _{t} &=&-\kappa \psi -\frac{1}{2}\phi _{xx}+\left( |\phi |^{2}+g|\psi
|^{2}\right) \phi ,  \label{GP1} \\
i\psi _{t} &=&-\kappa \phi -\frac{1}{2}\psi _{xx}+\left( |\psi |^{2}+g|\phi
|^{2}\right) \psi ,  \label{GP2}
\end{eqnarray}%
where $t$ is time, $x$ the spatial coordinate, $g$ the above-mentioned
relative strength of the cross-repulsion, and $\kappa $, which is defined to
be positive, is the coefficient of the linear mixing between components $%
\phi $ and $\psi $. If $\phi $ and $\psi $ represent two different hyperfine
states of the same atom, the linear mixing is induced by a resonant
electromagnetic (radiofrequency) field \cite{coupler,Kennedy}.

The same system of Eqs. (\ref{GP1}) and (\ref{GP2}), with $g=2$, time $t$
replaced by the propagation distance, $z$, and coordinate $x$ replaced by
the reduced time, $\tau =t-z/V_{\mathrm{gr}}$, is the pair of coupled NLS
equations governing the copropagation of two circular polarizations of
light, with group velocity $V_{\mathrm{gr}}$ of the carrier wave, in an
optical fiber with the self-defocusing cubic nonlinearity \cite{Agrawal}. In
that case, the linear mixing is induced by an elliptic deformation of the
originally circular fiber's core (in other words, by the birefringence of
the linearly polarized modes).

Stationary solutions to Eqs. (\ref{GP1}) and (\ref{GP2}) with chemical
potential $-k$ (or propagation constant $k$, in the case of the
optical-fiber model) are looked for as%
\begin{equation}
\phi \left( x,t\right) =e^{ikt}u(x),~\psi \left( x,t\right) =e^{ikt}v(x).
\label{k-GP}
\end{equation}%
The substitution of expressions (\ref{k-GP}) in Eqs. (\ref{GP1}) and (\ref%
{GP2}) yields equations%
\begin{eqnarray}
-ku+\kappa v &=&-\frac{1}{2}\frac{d^{2}u}{dx^{2}}+u^{3}+gv^{2}u,  \label{uxx}
\\
-kv+\kappa u &=&-\frac{1}{2}\frac{d^{2}v}{dx^{2}}+v^{3}+gu^{2}v.  \label{vxx}
\end{eqnarray}%
for real functions $u(x)$ and $v(x)$.

The norm of the stationary solution (\ref{k-GP}), which determines the total
number of atoms in the binary BEC, or the total beam power in the case of
the optical realization of Eqs. (\ref{GP1}) and (\ref{GP2}), is%
\begin{equation}
N=\int_{-\infty }^{+\infty }n(x)dx,~n(x)=\left\vert u(x)\right\vert
^{2}+\left\vert v(x)\right\vert ^{2}.  \label{N}
\end{equation}%
Similarly, the Hamiltonian of the coupled system (\ref{GP1}), (\ref{GP2}),
corresponding to the same stationary solution, is
\begin{equation}
H=\int_{-\infty }^{+\infty }h(x)dx,  \label{H}
\end{equation}
with the Hamiltonian density%
\begin{equation}
h=\frac{1}{2}\left[ \left( \frac{du}{dx}\right) ^{2}+\left( \frac{dv}{dx}%
\right) ^{2}\right] +\frac{1}{2}\left( u^{4}+v^{4}\right)
+gu^{2}v^{2}-2\kappa uv.  \label{h}
\end{equation}

In particular, the above-mentioned immiscibility condition $g>1$ (in the
case of $\kappa =0$) \cite{Mineev} immediately follows from Eq. (\ref{h}),
comparing the mixed and demixed states that occupy a large domain of size $L$%
. For the former state, with densities $u^{2}=v^{2}\equiv n/2$, which are
uniformly distributed in the entire domain, and for the latter state, which
has densities $\left( u^{2}=n,v^{2}=0\right) $ and $\left(
u^{2}=0,v^{2}=n\right) $ in two halves of the domain, separated by the DW,
so that both states share the same total norm, $N=nL$, the values of the
energy (Hamiltonian), as produced by Eqs. (\ref{H}) and (\ref{h}), are%
\begin{equation}
E_{\mathrm{mixed}}=\left( g+1\right) Ln^{2}/4,~E_{\mathrm{demixed}}=Ln^{2}/2
\label{mim}
\end{equation}%
(where the DW energy, which does not grow with the increase of $L$, is
neglected). The comparison of the energies immediately demonstrates $~E_{%
\mathrm{demixed}}<~E_{\mathrm{mixed}}$ at $g>1$, i.e., the immiscible state
provides the energy minimum in this case.

\section{Exact analytical results}

\subsection{SSB (spontaneous symmetry breaking) of CW states}

For CW ($x$-independent) states, the derivatives in Eqs. (\ref{uxx}) and (%
\ref{vxx}) are dropped, reducing them to a linearly coupled system of cubic
equations:%
\begin{eqnarray}
-ku+\kappa v &=&u^{3}+gv^{2}u,  \label{u} \\
-kv+\kappa u &=&v^{3}+gu^{2}v.  \label{v}
\end{eqnarray}%
A straightforward solution of the coupled equations (\ref{u}) and (\ref{v})
yields the symmetric state,%
\begin{equation}
u_{\mathrm{s}}=v_{\mathrm{s}}=\sqrt{\frac{\kappa -k}{g+1}},  \label{s}
\end{equation}%
with density%
\begin{equation}
n_{\mathrm{s}}=u_{\mathrm{s}}^{2}+v_{\mathrm{s}}^{2}=2\frac{\kappa -k}{g+1}.
\label{n_s}
\end{equation}%
and the antisymmetric state,%
\begin{equation}
u_{\mathrm{anti}}=-v_{\mathrm{anti}}=\sqrt{\frac{-k-\kappa }{g+1}}.
\label{antis}
\end{equation}%
These states exist in the regions of the chemical potential
\begin{equation}
\mathrm{symm:~}k<\kappa ;~\mathrm{antisymm:~}k<-\kappa   \label{kkappa}
\end{equation}%
(recall $\kappa >0$ is fixed by definition; in the absence of the linear
coupling, i.e., $\kappa =0$, Eq. (\ref{kkappa}) reduces to $k<0$). In fact,
the antisymmetric state is not relevant, because the corresponding
linear-coupling term $-2\kappa uv$ in the Hamiltonian density (\ref{h}) is
positive, unlike the negative one corresponding to the symmetric state,
hence the antisymmetric state tends to be unstable, as it cannot realize the
energy minimum.

SSB is represented by the exact asymmetric solution of the coupled equations
(\ref{u}) and (\ref{v}), which is obtained replacing the equations by their
sum and difference, and factorizing the resultant equations:
\begin{equation}
\left\{ u_{\mathrm{as}}^{2},v_{\mathrm{as}}^{2}\right\} =-\frac{k}{2}\pm
\sqrt{\frac{k^{2}}{4}-\frac{\kappa ^{2}}{\left( g-1\right) ^{2}}}.
\label{as}
\end{equation}%
The relative sign of the components of this solution is determined by the
relation%
\begin{equation}
u_{\mathrm{as}}v_{\mathrm{as}}=\frac{\kappa }{g-1}  \label{sign}
\end{equation}%
(which is compatible with Eq. (\ref{as})), i.e., the signs of both
components are identical in the case of interest, $g>1$, where the
immiscibility is possible. Thus, unlike the SSB effect, spontaneous breaking
of the antisymmetry is not possible.

The asymmetric solution (\ref{as}) exists, for given $\kappa >0$, in the
following region of the chemical potential ($-k$):%
\begin{equation}
-k>-k_{\mathrm{thr}}\equiv \frac{2\kappa }{g-1},  \label{kthr}
\end{equation}%
cf. Eq. (\ref{kkappa}), or, in terms of the density,%
\begin{equation}
n_{\mathrm{as}}\equiv u_{\mathrm{as}}^{2}+v_{\mathrm{as}}^{2}=-k>n_{\mathrm{%
thr}}\equiv \frac{2\kappa }{g-1},  \label{nthr}
\end{equation}%
where $k_{\mathrm{thr}}$ is the value of $k<0$ at which the square root
vanishes in Eq. (\ref{as}). With the increase of the density of the
symmetric state [see Eq. (\ref{n_s})], the SSB bifurcation takes place at
the threshold value (\ref{nthr}), at which the symmetric solution (\ref{s})
is%
\begin{equation}
\left( u_{\mathrm{s}}^{2}\right) _{\mathrm{thr}}=\left( v_{\mathrm{s}%
}^{2}\right) _{\mathrm{thr}}=\frac{\kappa }{g-1},  \label{thr}
\end{equation}%
cf. Eq. (\ref{thr}).

It is natural to expect that the transition from the symmetric state to the
asymmetric one leads to the decrease of the state's energy. Indeed, the
substitution of the asymmetric solution (\ref{as}), (\ref{sign}) in the
Hamiltonian density (\ref{h}) yields the following expression, as a function
of the density:%
\begin{equation}
h_{\mathrm{as}}=\frac{n^{2}}{2}-\frac{\kappa ^{2}}{g-1}.  \label{h_as}
\end{equation}%
On the other hand, the substitution of the symmetric CW state (\ref{s}) in
expression (\ref{h}), taking into regard expression (\ref{n_s}) for the
respective density, yields%
\begin{equation}
h_{\mathrm{s}}=\frac{(g+1)}{4}n^{2}-\kappa n.  \label{h_s2}
\end{equation}%
Then, it follows from Eqs. (\ref{h_as}) and (\ref{h_s2}) that, in the case
of $n>n_{\mathrm{thr}}$, when the symmetric and asymmetric CW states coexist
for the same density, the difference between their energy densities is%
\begin{equation}
h_{\mathrm{s}}-h_{\mathrm{as}}=\frac{g-1}{4}\left( n-n_{\mathrm{thr}}\right)
^{2}>0.  \label{h-h}
\end{equation}%
Thus, the SSB transition from the symmetric state to the asymmetric one
indeed leads to the decrease of the energy.

\subsection{Modulational stability of the symmetric CW state at the
SSB-bifurcation point}

The CW states found above are physically relevant ones if they are not
subject to MI. The modulational stability can be explored in an analytical
form for the symmetric CW state, while the analysis is too cumbersome for
the asymmetric one. The most relevant result is the one for the symmetric CW
at the SSB-bifurcation point (\ref{kthr}), (\ref{thr}), as the verification
of the modulational stability at this point implies that the asymmetric CW
inherits the stability.

To perform the analysis, solutions to Eqs. (\ref{GP1}) and (\ref{GP2}) are
looked for in the Madelung form:%
\begin{equation}
\phi \left( x,t\right) =a\left( x,t\right) \exp \left( i\alpha \left(
x,t\right) \right) ,\psi \left( x,t\right) =b\left( x,t\right) \exp \left(
i\beta \left( x,t\right) \right) ,  \label{Madelung}
\end{equation}%
where $a\left( x,t\right) ,b\left( x,t\right) $ and $\alpha \left(
x,t\right) ,\beta \left( x,t\right) $ are real amplitudes and phases of wave
functions $\phi $ and $\psi $. The substitution of expressions (\ref%
{Madelung}) leads to a system of four real equations:%
\begin{gather}
a_{t}+\frac{1}{2}a\alpha _{xx}+a_{x}\alpha _{x}-\kappa b\sin =0,  \label{a}
\\
b_{t}+\frac{1}{2}b\beta _{xx}+b_{x}\beta _{x}+\kappa a\sin \left( \alpha
-\beta \right) =0,  \label{b} \\
-a\alpha _{t}+\frac{1}{2}a_{xx}-\frac{1}{2}a\alpha _{x}^{2}+\kappa b\cos
\left( \alpha -\beta \right) -a^{3}-2b^{2}a=0,  \label{alpha} \\
-b\beta _{t}+\frac{1}{2}b_{xx}-\frac{1}{2}b\beta _{x}^{2}+\kappa a\cos
\left( \alpha -\beta \right) -b^{3}-2ab=0.  \label{beta}
\end{gather}%
According to Eqs. (\ref{kthr}) and (\ref{thr}), the symmetric CW (\ref{s})
at the bifurcation point corresponds to the solution of equations (\ref{a})-(%
\ref{beta}) in the form of $\alpha =\beta =-2\kappa (g-1)^{-1}t$, $a_{%
\mathrm{s}}=b_{\mathrm{s}}=\sqrt{\kappa /(g-1)}$. For the stability\
analysis, perturbed solutions are looked for as%
\begin{eqnarray}
\left\{ a\left( x,t\right) ,b\left( x,t\right) \right\} &=&\sqrt{\kappa
/(g-1)}+\left\{ a_{1},b_{1}\right\} \exp \left( \gamma t-iqx\right) ,
\label{ab} \\
\left\{ \alpha \left( x,t\right) ,\beta \left( x,t\right) \right\}
&=&-2\kappa (g-1)^{-1}t+\left\{ \alpha _{1},\beta _{1}\right\} \exp \left(
\gamma t-iqx\right) ,~  \label{alphabeta}
\end{eqnarray}%
where $a_{1},b_{1},\alpha _{1},\beta _{1}$ are amplitudes of infinitesimal
perturbations, $\gamma $ is the instability growth rate, and $q$ is an
arbitrary real wavenumber of the perturbation. MI takes place if one has Re$%
\left( \gamma (q)\right) >0$ at any real value of $q$.

The substitution of the perturbed solution (\ref{ab}), (\ref{alphabeta}) in
Eqs. (\ref{a})-(\ref{beta}) and linearization with respect to the
infinitesimal perturbations leads to the dispersion equation, which
determines $\gamma $ as a function of $q$:%
\begin{equation}
16(g-1)\gamma ^{4}+8q^{2}\left[ 4g\kappa +(g-1)q^{2}\right] \allowbreak
\gamma ^{2}+q^{4}\left[ (g-1)q^{4}+8g\kappa q^{2}+16\kappa ^{2}(g+1)\right]
=0.  \label{disp}
\end{equation}%
All roots of the biquadratic equation (\ref{disp}) are pure imaginary ones:%
\begin{equation}
\gamma _{1,2}=\pm iq\sqrt{\kappa +\frac{q^{2}}{4}},~\gamma _{3,4}=\pm iq%
\sqrt{\kappa \frac{g+1}{g-1}+\frac{q^{2}}{4}},
\end{equation}%
hence there is no MI at the SSB-bifurcation point.

\subsection{The exact solution for the symmetry-breaking dark soliton (DS)
in the case of $g=3$}

\bigskip The symmetric and antisymmetric CW states, given by Eqs. (\ref{s})
and (\ref{antis}), respectively, support symmetric and antisymmetric
two-component DSs:%
\begin{equation}
v_{\mathrm{DS}}(x)=u_{\mathrm{DS}}(x)=\sqrt{\frac{-k+\kappa }{g+1}}\tanh
\left( \sqrt{\kappa -k}x\right) ;  \label{DS}
\end{equation}%
\begin{equation}
v_{\mathrm{DS}}(x)=-u_{\mathrm{DS}}(x)=\sqrt{\frac{-k-\kappa }{g+1}}\tanh
\left( \sqrt{-k-\kappa }x\right) .  \label{anti}
\end{equation}%
Note that the symmetric and antisymmetric DSs (\ref{DS}) and (\ref{anti})
are transformed into each other by the substitution,
\begin{equation}
u\rightarrow u,v\rightarrow -v,\kappa \rightarrow -\kappa ,x\rightarrow -x,
\label{subst}
\end{equation}%
with respect to which the systems of Eqs. (\ref{GP1}), (\ref{GP2}) and (\ref%
{uxx}), (\ref{vxx}) are invariant.

On the other hand, it has been known since long ago \cite{1990} that the
system of coupled real NLS equations (\ref{uxx}), (\ref{vxx}) with the
special value $g=3$ of the inter-component interaction coefficient and no
linear coupling ($\kappa =0$) admits an exact DW solution. Then, the exact
solution with $g=3$ was extended to the system of Eqs. (\ref{uxx}), (\ref%
{vxx}) with $\kappa \neq 0$ \cite{PLA}:%
\begin{equation}
\left\{
\begin{array}{c}
u_{\mathrm{DW}}(x) \\
v_{\mathrm{DW}}(x)%
\end{array}%
\right\} =\frac{1}{2}\left\{
\begin{array}{c}
\sqrt{-k+\kappa }+\sqrt{-k-\kappa }\tanh \left( \sqrt{-k-\kappa }x\right) \\
\sqrt{-k+\kappa }-\sqrt{-k-\kappa }\tanh \left( \sqrt{-k-\kappa }x\right)%
\end{array}%
\right\} ,  \label{exact2}
\end{equation}%
which satisfies the boundary conditions%
\begin{equation}
u\left( x=+\infty \right) =v\left( x=-\infty \right) \equiv v_{\mathrm{as}%
}(g=3),v\left( x=+\infty \right) =u\left( x=-\infty \right) \equiv u_{%
\mathrm{as}}(g=3),  \label{twisted}
\end{equation}%
with $u_{\mathrm{as}}(g=3)$ and $v_{\mathrm{as}}(g=3)$ given by Eqs. (\ref%
{as}) and (\ref{sign}) with $g=3$. Thus, the DW state (\ref{exact2})
separates two asymmetric CW states which are mirror images of each other.

A new result reported here is the following exact DS solution of the system
of Eqs. (\ref{uxx}) and (\ref{vxx}), which is actually produced by the
application of substitution (\ref{subst}) to the DW (\ref{exact2}):%
\begin{equation}
\left\{
\begin{array}{c}
u_{\mathrm{DS}}(x) \\
v_{\mathrm{DS}}(x)%
\end{array}%
\right\} =\frac{1}{2}\left\{
\begin{array}{c}
\sqrt{-k-\kappa }+\sqrt{-k+\kappa }\tanh \left( \sqrt{-k+\kappa }x\right) \\
-\sqrt{-k-\kappa }+\sqrt{-k+\kappa }\tanh \left( \sqrt{-k+\kappa }x\right)%
\end{array}%
\right\} ,  \label{dark}
\end{equation}%
which is subject to the boundary conditions%
\begin{equation}
u\left( x=+\infty \right) =-v\left( x=-\infty \right) \equiv u_{\mathrm{as}%
}(g=3),v\left( x=+\infty \right) =-u\left( x=-\infty \right) \equiv v_{%
\mathrm{as}}(g=3),  \label{bc}
\end{equation}%
cf. Eq. (\ref{twisted}). The relation of the present DS solution to the
background asymmetric CW states, fixed as per Eq. (\ref{bc}), implies that
the DS is a result of the SSB occurring with the symmetric two-component DS (%
\ref{DS}). On the other hand, the DS solution (\ref{dark}), being subject to
the boundary conditions (\ref{bc}), demonstrates its inversion symmetry,
expressed by relation (\ref{phipsi}) [in other words, by its counterpart $v_{%
\mathrm{DS}}(-x)=-u_{\mathrm{DS}}(x)$].

Unlike the DW solution (\ref{exact2}), which keeps positive values of both
components at all $x$, the two components of DS (\ref{dark}) have
zero-crossing [$u\left( x_{\mathrm{ZC}}^{(u)}\right) =0$, $v\left( x_{%
\mathrm{ZC}}^{(v)}\right) =0$] points, the presence of which is the hallmark
of DS states. These points are determined by relations%
\begin{equation}
\tanh \left( \sqrt{-k+\kappa }x_{\mathrm{ZC}}^{(u,v)}\right) =\left(
-,+\right) \sqrt{\frac{-k-\kappa }{-k+\kappa }},  \label{zero-crossing}
\end{equation}%
Both the DW and DS solutions (\ref{exact2}) and (\ref{dark}) exist in the
region of $k<-\kappa $, which coincides with the existence condition (\ref%
{kthr}) for the asymmetric CW state (\ref{as}), with $g=3$.

\subsection{The system with the P\"{o}schl-Teller (PT) potential}

To construct an exact DS solution with an asymmetric CW background at $g\neq
3$, one can add the PT potential to the underlying system of Eqs. (\ref{GP1}%
)-(\ref{GP2}):%
\begin{eqnarray}
i\frac{\partial \phi }{\partial t} &=&-\frac{1}{2}\frac{\partial ^{2}\phi }{%
\partial x^{2}}+\left( |\phi |^{2}+g|\psi |^{2}\right) \phi +\frac{W}{\cosh
^{2}(\alpha x)}\phi -\kappa \psi ,  \label{PT+} \\
i\frac{\partial \psi }{\partial t} &=&-\frac{1}{2}\frac{\partial ^{2}\psi }{%
\partial x^{2}}+\left( |\psi |^{2}+g|\phi |^{2}\right) \psi +\frac{W}{\cosh
^{2}(\alpha x)}\psi -\kappa \phi ,  \label{PT-}
\end{eqnarray}%
with specially selected values of the potential's strength and width:%
\begin{equation}
W=\frac{3-g}{g-1}\frac{g-1+2\kappa }{4},~\alpha =\sqrt{\frac{g-1+2\kappa }{2}%
}.  \label{Walpha}
\end{equation}%
This potential can be readily implemented in the experimental setting for
BEC with the help of an appropriately shaped laser beam illuminating the
condensate \cite{Hulet,Cornish}. Note that the PT potential is attractive or
repulsive [with Eq. (\ref{Walpha}) yielding $W<0$ or $W>0$] in the case of $%
g>3$ or $g<3$, respectively.

The system of Eqs. (\ref{PT+}) and (\ref{PT-}) with the PT potential subject
to conditions (\ref{Walpha}) admits the exact DS solution:%
\begin{equation}
\left\{
\begin{array}{c}
\phi (x) \\
\psi (x)%
\end{array}%
\right\} =\frac{1}{2}e^{-it}\left\{
\begin{array}{c}
A+B\,\mathrm{tanh}(\alpha \,x) \\
-A+B~\mathrm{tanh}(\alpha \,x)%
\end{array}%
\right\} ,  \label{exact4}
\end{equation}%
with amplitudes
\begin{equation}
A=\sqrt{\frac{g-1-2\kappa }{g-1}},~B=\sqrt{\frac{g-1+2\kappa }{g-1}},
\label{AB2}
\end{equation}%
cf. the above solution (\ref{dark}). Factor $e^{-it}$ in expression (\ref%
{exact4}) implies that the chemical potential is here fixed as $-k=1$ by
means of scaling. It is easy to see that, with respect to the adopted
scaling $k=-1$, the CW background maintaining the DS produced by expressions
(\ref{exact4}) and (\ref{AB2}) is exactly given by Eq. (\ref{as}). In
particular, the present DS solution exists at $g-1>2\kappa $, which is
exactly the same as condition (\ref{kthr}) (with $k=-1$), under which the
asymmetric CW solution (\ref{as}) exists.

Due to the inequality $B>A$, following from Eq. (\ref{AB2}), the DS solution
(\ref{exact4}) demonstrates that both DS components have zero-crossing
points determined by relations%
\begin{equation}
\tanh \left( \alpha x_{\mathrm{ZC}}^{(\phi ,\psi )}\right) =\left(
-,+\right) \frac{A}{B},  \label{ZC}
\end{equation}
cf. Eq. (\ref{zero-crossing}). As well as the exact DS solution (\ref{dark}%
), the one given by Eqs. (\ref{exact4}) and (\ref{AB2}) is also subject to
the inversion symmetry (\ref{phipsi}). However, unlike the stable DS (\ref%
{dark}), the one given by Eqs. (\ref{PT+})-(\ref{AB2}) is unstable, as
indicated below.

\section{Numerical and variational results for DSs (dark solitons) in 1D and
vortices in 2D}

\subsection{Symmetry-breaking 1D dark solitons: numerical results}

While the SSB phenomenology for the CW solutions of the coupled GP equations
(\ref{GP1}) and (\ref{GP2}) is fully described by the simple analytical
solution provided by Eqs. (\ref{as}) and (\ref{sign}), the solution for the
DS supported by the asymmetric DW background is available in the analytical
form, given by Eq. (\ref{dark}), only in the special case of $g=3$ [and
also, for $g\neq 3$, in the presence of the specially designed PT potential
with parameters (\ref{Walpha})]. In this subsection we aim to produce
generic symmetry-breaking DS solutions, for $g\neq 3$, in the numerical form
and verify their stability.

The stationary DS solutions were obtained by means of the imaginary-time
propagation method \cite{Tosi,Bao}, while their stability was tested by
simulations of the perturbed evolution in real time. The numerical solutions
were actually constructed for a pair of DSs, placed at diametrically
opposite positions in a domain of a large size $L$, subject to periodic
boundary conditions, as this setting makes it possible to eliminate
artifacts induced by boundary conditions of other types (Dirichlet's or
Neumann's, fixed at edges of a non-circular solution domain).

\begin{figure}[h]
\begin{center}
\includegraphics[height=3.5cm]{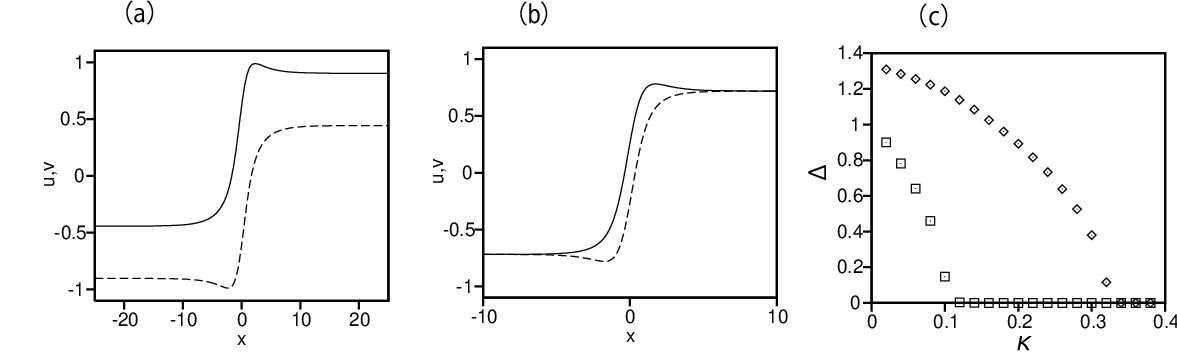}
\end{center}
\caption{(a) and (b): Profiles of components $u(x)$ and $v(x)$ in the\
numerically constructed stable symmetry-breaking DSs which connect,
respectively, the partly immiscible or fully miscible CW states at $x=\pm
\infty $. The corresponding parameters are given by Eq. (\protect\ref{1a})
for (a) and (\protect\ref{1b}) for (b). In both cases, the total norm is $%
N=100$, and the solutions are constructed, as DS pairs, on the circumference
of perimeter $L=100$ (here, only one DS is displayed, in the interval of $%
-L/4<x<+L/4$). (c) The outer and inner immiscibilities in the DSs, with
fixed $N=100$, $L=100$, and $g=1.2$, are characterized by dependences of $%
\Delta \equiv |u(x)-v(x)|$ on the linear coupling $\protect\kappa $ at $%
x=-L/4$ (squares) and $x=0$ (rhombuses).}
\label{f1}
\end{figure}

In the case of the chemical potential $-k$ subject to condition (\ref{kthr}%
), when the stable CW state is the asymmetric one (\ref{as}), the
numerically found stable symmetry-breaking DSs connect these states,
satisfying the boundary conditions (\ref{bc}), with $g\neq 3$. An example is
shown in Fig. \ref{f1}(a), for
\begin{equation}
g=1.2,\kappa =0.06,k=-1.01  \label{1a}
\end{equation}%
[note that these values indeed satisfy condition (\ref{kthr})]. In this and
all other plots for 1D DSs, the total norm of the solution, constructed on
the circumference of perimeter $L=100$, is \ $\int_{-L/2}^{+L/2}\left[
|u(x)|^{2}+|v(x)|^{2}\right] dx=100$ [cf. definition (\ref{N}) for the
infinite domain]. A peculiarity of the exact DS solution (\ref{dark}),
available at $g=3$, is that the difference of its two components is a
constant, which does not depend on coordinate $x$: $u_{\mathrm{DS}%
}(x;g=3)-v_{\mathrm{DS}}(x;g=3)=\sqrt{-k-\kappa }$ at any $x$. On the
contrary, at $g\neq 3$ numerically found DS solitons have $u(x)-v(x)$
depending on $x$ [in particular, this is the case for the DS shown in Fig. %
\ref{f1}(a)].

A new finding produced by the numerical solutions is that symmetry-breaking
DS solitons, with splitting between $u(x)$ and $v(x)$ (i.e., partial
immiscibility of the two components) occurring in the core of the DS, exist
as well in the case when condition (\ref{kthr}) \emph{does not hold}, i.e.,
the asymmetric CW states \emph{do not exist} (in other words, the CW state
remains \emph{fully miscible}). An example of such a stable DS, supported by
the symmetric CW background (\ref{s}), with the partial immiscibility
observed in the DS's core, is presented in Fig. \ref{f1}(b), for parameters%
\begin{equation}
g=1.2,\kappa =0.28,k=-0.859.  \label{1b}
\end{equation}%

The\emph{\ }above-mentioned \textit{outer} and \textit{inner}
immiscibilities, i.e., the existence of the DS states supported by the
asymmetric CW background, as determined by condition (\ref{kthr}), and, on
the other hand, the splitting of $u(x)$ and $v(x)$ in the core of the DS
supported by the miscible (symmetric) CW background, are quantified in Fig. %
\ref{f1}(c) by the respective splittings, i.e., $\Delta \equiv |u(x)-v(x)|$
at $x=L/4$ (far from the DS core) and at $x=0$ (at the DS center), plotted
vs. the linear-coupling constant $\kappa $. It is seen that the outer
immiscibility occurs at $\kappa <0.11$ [in agreement with condition (\ref%
{kthr})], while the inner immiscibility persists at essentially larger
values of the coupling, \textit{viz}., at $\kappa <0.33$. The latter effect
is qualitatively explained by the fact that the opposite signs of $u(x)$ and
$v(x)$ in the DS core make the corresponding linear-mixing term $-2\kappa uv$
in the energy density (\ref{h}) positive, which is an obstacle for mixing,
unlike the negative mixing term in the energy density for the CW state.
Below, a more accurate explanation of the inner immiscibility in the case of
the symmetric CW background is proposed, based on VA.
\begin{figure}[h]
\begin{center}
\includegraphics[height=3.5cm]{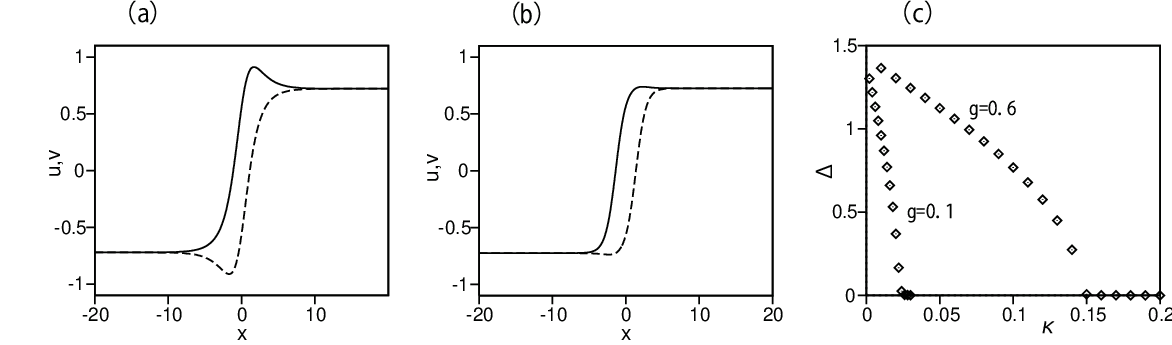}
\end{center}
\caption{(a) and (b): The same as in Fig. \protect\ref{f1}(b), but for $g=1.0
$, $\protect\kappa =0.1$, $k=-0.925$, and $g=0.1$, $\protect\kappa =0.006$, $%
k=-0.573$. (c) The measure of the core's immiscibility, $\Delta \equiv
\left\vert u(x=0)-v(x=0)\right\vert $, for the DSs with the inner shift
between the components, vs. the linear coupling, $\protect\kappa $, at $g=0.1
$ and $0.6$.}
\label{f2}
\end{figure}
\begin{figure}[h]
\begin{center}
\includegraphics[height=3.5cm]{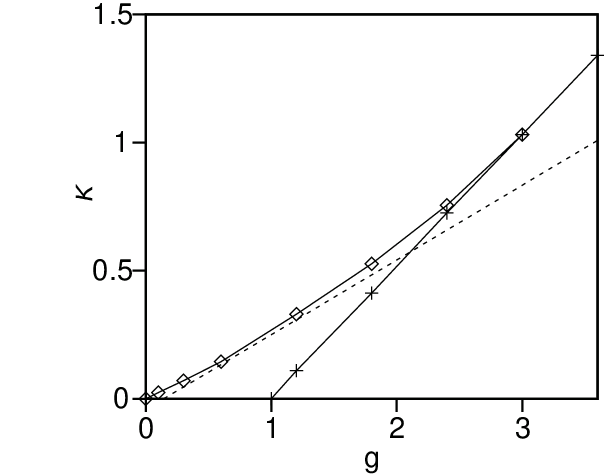}
\end{center}
\caption{In the parameter plane of the inter-component repulsion coefficient
($g$) and linear-mixing coefficient ($\protect\kappa $), the outer and inner
immiscibilities take place at $\protect\kappa <\protect\kappa _{\mathrm{c}%
}(g)$, i.e., beneath the chains of crosses and squares, respectively. The
solid line with crosses is given by Eq. (\protect\ref{kappa_c}) with density
$n=1$. The two critical lines merge at $g\geq 3$, where there is no
difference between the outer and inner immiscibility. The dashed line is the
VA prediction for the boundary of the inner immiscibility, as given by Eq. (%
\protect\ref{var}) with $n=1$.}
\label{f3}
\end{figure}

Furthermore, the immiscibility may occur in the DS core even in the case of $%
g\leq 1$, when the SSB (immiscibility) does not take place for the CW states
even in the absence of the linear coupling. Examples are displayed in Figs. %
\ref{f2}(a) and (b), for $g=1.0$, $\kappa =0.1$, $k=-0.925$ and $g=0.1$, $%
\kappa =0.006$, $k=-0.573$, respectively (note the similarity of the DS
profiles with the inner immiscibility in Figs. \ref{f1}(b) and \ref{f2}(a)).
Figure \ref{f2}(c) summarizes these results by plotting $\Delta \equiv
\left\vert u(x=0)-v(x=0)\right\vert $ vs. $\kappa $ at $g=0.1$ and $0.6$.
The inner immiscibility in the DS core takes place, accordingly, at $\kappa
<0.025$ for $g=0.1$ and $\kappa <0.145$ for $g=0.6$. Naturally, the largest
value of $\kappa $ admitting the inner immiscibility increases with the
increase of $g$, as the linear coupling ($\kappa $) and nonlinear repulsion (%
$g)$ of the two components are competing factors. It is natural too that no
inner immiscibility is observed for $g=0$.

The numerically found boundaries of the outer and inner immiscibility in the
parameter plane of $\left( g,\kappa \right) $ are plotted in Fig. \ref{f3}
by the dashed and solid lines, respectively. With fixed $N=L=100$, both
boundaries correspond to the CW background with density $n=N/L=1$. The
substitution of this value in Eq. (\ref{nthr}) yields the threshold
(critical) value of the linear coupling which determines the onset of SSB in
the CW state, i.e., the boundary of the outer immiscibility,
\begin{equation}
\kappa _{\mathrm{c}}=\left( g-1\right) n/2,  \label{kappa_c}
\end{equation}%
which exactly corresponds to the dashed line in Fig. \ref{f3}. The two
boundaries merge in Fig. \ref{f3} at $g=3$,\ where the DS is given by the
exact solution (\ref{dark}), there being no distinction between the outer
and inner immiscibility in the DS states at $g\geq 3$.

The stability of all DSs featuring the inner-only and outer immiscibilities,
including the exact solution given by Eq. (\ref{dark}) for $g=3$, has been
verified by simulations of their perturbed evolution. On the other hand, it
was found that all exact symmetry-breaking DS solutions (\ref{exact4}),
attached to the PT potential, are unstable, for $g<3$ and $g>3$ alike (not
shown here in detail). In particular, in the case of $g>$ $3$, when the PT
potential (\ref{Walpha}) is attractive, the instability is explained by the
general fact that DSs are \textit{repelled} by attractive local potentials
(see, e.g., Ref. \cite{Ricardo}). In the case of $g<3$, the effective
\textit{attraction} of the zero-crossing points (\ref{ZC}) to the local
\textit{repulsive} potential also implies an instability. It is relevant to
mention that, along with the DSs, the PT potential also admits exact
solutions of the DW type. Additional analysis (not presented here)
demonstrates that the DW pinned to the PT potential is stable at $g<3$ and
unstable $g>3$ (a similar result for the DWs was recently obtained in Ref.
\cite{BBB}).

\subsection{The variational approximation (VA) for the DS with the inner
immiscibility}

While the onset of the outer immiscibility in the DSs is readily explained
by the exact solution (\ref{as}) for the symmetry-breaking CW background
states, the inner immiscibility, which is exhibited, in Figs. \ref{f1}(b)
and \ref{f2}(a), by the DS cores supported by the symmetric CW background,
can be explained, in a crude analytical approximation, by means of VA (see
Ref. \cite{Progress} for a review). To this end, we use the Lagrangian of
the system of Eqs. (\ref{uxx}) and (\ref{vxx}), defined on the circumference
of perimeter $L$:
\begin{equation}
\mathcal{L}=\int_{-L/2}^{+L/2}\left\{ \frac{1}{4}\left[ \left( \frac{%
\partial u}{\partial x}\right) ^{2}+\left( \frac{\partial v}{\partial x}%
\right) ^{2}\right] +\frac{k}{2}\left( u^{2}+v^{2}\right) +\frac{1}{4}\left(
u^{4}+v^{4}\right) +\frac{g}{2}u^{2}v^{2}-\kappa uv\right\} dx.  \label{en}
\end{equation}%
A natural ansatz for a DS with the $u$ and $v$ components separated
(demixed) inside the core and merged (mixed) in the CW background is adopted
as%
\begin{equation}
u=U\left[ \tanh (\xi x)+\frac{a}{\cosh (\xi x)}\right] ,\;\;v=U\left[ \tanh
(\xi x)-\frac{a}{\cosh (\xi x)}\right] ,  \label{ans}
\end{equation}%
with coefficients $U$ and $\xi $ taken as per the symmetric DS solution (\ref%
{DS}), with density $n$ defined as per Eq. (\ref{n_s}):%
\begin{equation}
U=\sqrt{\frac{\kappa -k}{1+g}}\equiv \sqrt{\frac{n}{2}},\;\;\xi =\sqrt{%
\kappa -k},  \label{A}
\end{equation}%
while $a$ is a free variational parameter. The substitution of ansatz (\ref%
{ans}) in Lagrangian (\ref{en}) yields%
\begin{gather}
\mathcal{L}=\frac{U^{2}}{3\xi }\left( \xi ^{2}+6U^{2}+6k+6\kappa
-2gU^{2}a^{2}\right) +\frac{2}{3\xi }(1+g)U^{4}a^{4}  \notag \\
+L\left[ \left( k-\kappa \right) \frac{n}{2}+\left( g+1\right) \frac{n^{2}}{8%
}\right] ,  \label{L}
\end{gather}%
where the last (bulk) term, $\sim L$, is an $a$-independent contribution
from the CW background. Further, the chemical potential $-k$ in Eq. (\ref{L}%
) is replaced by expression $-k=(1+g)n/2-\kappa $, according to Eq. (\ref{A}%
). Then, the variational (Euler-Lagrange) equation is produced by Lagrangian
(\ref{L}) as $\partial \mathcal{L}/\partial \left( a^{2}\right) =0$. After a
simple algebra, the latter equation predicts the following critical
(threshold) value of the linear coupling at which the inner immiscibility
commences:%
\begin{equation}
\kappa _{\mathrm{c}}=\left( 7g-1\right) n/24,  \label{var}
\end{equation}%
cf. its counterpart (\ref{kappa_c}), which is the exact result for the CW
states [negative value produced by Eq. (\ref{var}) at $g<1/7$ is an artifact
of the VA].

\section{Vortex states with the inter-component shift in the two-dimensional
binary BEC}

The system of coupled GP equations for the 2D binary BECs is written as a
straightforward extension of the 1D system of Eqs. (\ref{GP1}) and (\ref{GP2}%
):
\begin{eqnarray}
i\phi _{t} &=&-\kappa \psi -\frac{1}{2}\left( \phi _{xx}+\phi _{yy}\right)
+\left( |\phi |^{2}+g|\psi |^{2}\right) \phi ,  \label{GP3} \\
i\psi _{t} &=&-\kappa \phi -\frac{1}{2}\left( \psi _{xx}+\psi _{yy}\right)
+\left( |\psi |^{2}+g|\phi |^{2}\right) \psi .  \label{GP4}
\end{eqnarray}%
Stationary 2D states are looked for as%
\begin{equation}
\phi \left( x,y,t\right) =e^{ikt}u(x,y),~\psi \left( x,y,t\right)
=e^{ikt}v(x,y),  \label{k-GP2}
\end{equation}%
with complex functions $u(x,y)$ and $v(x,y)$ satisfying equations%
\begin{eqnarray}
-ku+\kappa v &=&-\frac{1}{2}\left( \frac{d^{2}}{dx^{2}}+\frac{d^{2}}{dy^{2}}%
\right) u+\left( |u|^{2}+g|v|^{2}\right) u,  \label{2Du} \\
-kv+\kappa u &=&-\frac{1}{2}\left( \frac{d^{2}}{dx^{2}}+\frac{d^{2}}{dy^{2}}%
\right) v+\left( |v|^{2}+g|u|^{2}\right) v.  \label{2Dv}
\end{eqnarray}

Vortex states, supported by the CW background, are 2D counterparts of the 1D
DSs. In terms of\ phases $\alpha \left( x,y\right) $ and $\beta \left(
x,y\right) $ of the complex fields $u\left( x,y\right) $ and $v\left(
x,y\right) $ [cf. Eq. (\ref{Madelung})], integer vorticity $S$ (alias the
winding number, or topological charge) is defined as usual \cite{vortex}: $%
S=\Delta \alpha /(2\pi )=\Delta \beta /(2\pi )$, where $\Delta \alpha $ and $%
\Delta \beta $ are phase circulations along trajectories enclosing the
vortex core. The 2D states with $S=0$ amount to the CW solution.

The systems of Eqs. (\ref{GP3}), (\ref{GP4}) and (\ref{2Du}), (\ref{2Dv})
was solved in the 2D domain $-L/2<x,y<+L/2$, with periodic boundary
conditions imposed in both directions. Accordingly, a vortex-antivortex pair
is created under the periodic boundary condition, while here we display the
vortex centered at the origin. The solutions are characterized by the 2D
norm,%
\begin{equation}
N_{\mathrm{2D}}=\int_{-L/2}^{+L/2}dx\int_{-L/2}^{+L/2}dy\left[ \left\vert
\phi (x,y)\right\vert ^{2}+\left\vert \psi (x,y)\right\vert ^{2}\right] .
\label{N2D}
\end{equation}

\begin{figure}[h]
\begin{center}
\includegraphics[height=3.5cm]{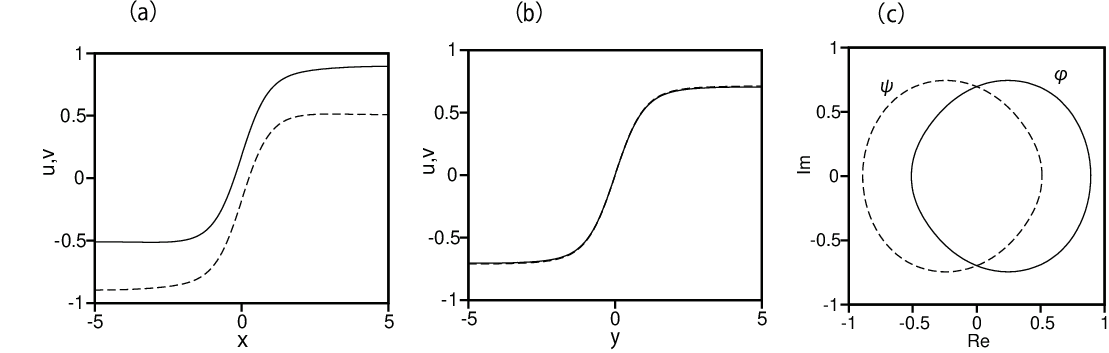}
\end{center}
\caption{(a) The cross-sections of \textrm{Re}$\left( u(x,y)\right) $ and
\textrm{Re}$\left( v(x,y)\right) $ (solid and dashed lines, respectively) in
the numerically found stable vortex state with $S=1$, maintained by the
asymmetric CW background, along $y=0$. (b) The cross sections of \textrm{Im}$%
\left( u(x,y)\right) $ and \textrm{Im}$\left( v(x,y)\right) $ (solid and
dashed lines, respectively) along $x=0$. (c) $(\mathrm{Re~}\protect\phi ,%
\mathrm{Im~}\protect\phi )$ and $(\mathrm{Re~}\protect\psi ,\mathrm{Im~}%
\protect\psi )$ (solid and dashed loops, respectively) along the
circumference of radius $r=4$. The parameters are given by Eq. (\protect\ref%
{2Dparam}).}
\label{f5}
\end{figure}

The symmetry-breaking CW states in the 2D case are the same as given by Eq.
(\ref{as}), and are equally stable. Stable vortex modes supported by such
background states demonstrate a specific 2D feature in the form of a shift
between their $u$ and $v$ components, as shown in Fig. \ref{f5}, for $S=1$
and parameters%
\begin{equation}
g=2,\kappa =0.45,k=-1.1,N=400,L=20  \label{2Dparam}
\end{equation}%
[note that these parameters satisfy condition (\ref{kthr}) which maintains
the asymmetric CW states]. The shift between the two components of the
vortex mode is a direct counterpart of the separation between the components
in the 1D DSs maintained by the asymmetric background. In the example
displayed in Fig. \ref{f5}, the vortical cores in the coupled components are
mutually shifted in the $x$ direction, thus breaking the cores' isotropy.
The direction of the shift may be arbitrary, selected by initial conditions.

\begin{figure}[h]
\begin{center}
\includegraphics[height=3.5cm]{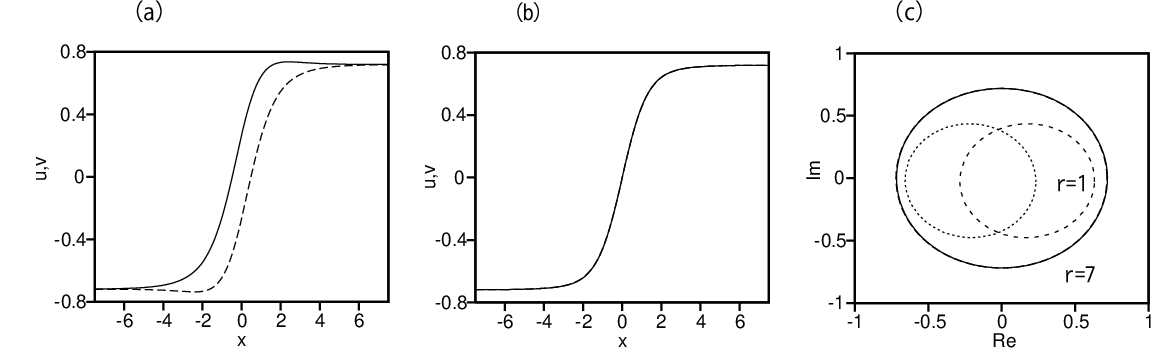}
\end{center}
\caption{Panels (a) and (b) display the same as in Figs. \protect\ref{f5}%
(a,b), but for parameters (\protect\ref{2Dparam2}). Panel (c) is similar to
Fig. \protect\ref{f6}(c), with the difference that the parameters are taken
as per Eqs. (\protect\ref{2Dparam2}), while $(\mathrm{Re~}\protect\phi ,%
\mathrm{Im~}\protect\phi )$ and $(\mathrm{Re~}\protect\psi ,\mathrm{Im~}%
\protect\psi )$ are plotted along two circumferences, with radii $r=1$ and $%
7 $.}
\label{f6}
\end{figure}

The consideration of the 1D\ system presented above has produced, in
addition to the DSs with the outer immiscibility, supported by the
symmetry-breaking CW background [see Fig. \ref{f1}(a)], also DS modes
featuring the inner immiscibility in their cores, while the CW background
remains symmetric [see Fig. \ref{f1}(b) and \ref{f2}(a); recall that such
modes populate the area between the solid and dashed lines in Fig. \ref{f3}%
]. Similarly, the 2D system is capable to produce stable vortices maintained
by the symmetric CW background, with the inner shift between their vortical
cores. An example is displayed in Fig. \ref{f6}(a), for\ $S=1$ and parameters%
\begin{equation}
g=0.8,\kappa =0.08,k=-0.865,N=900,L=30.  \label{2Dparam2}
\end{equation}%
The systematic analysis demonstrates that, for $g=0.8$ and $N=900,L=30$,
i.e.,\ with the background density $n\approx 1$, the shift (splitting)
between the vortical cores of the two components persists at $\kappa \leq
\kappa _{\mathrm{c}}\approx 0.095$ [in particular, $\kappa =0.08$ in Eq. (%
\ref{2Dparam2}) satisfies the latter condition]; for comparison, it is
relevant to mention that the critical value of $\kappa $ for supporting the
inner immiscibility in the 1D DS at $g=0.8$ and $n=1$ is much larger, $%
\kappa _{\mathrm{c}}\approx 0.205$. In particular, Fig. \ref{f6}(c) clearly
demonstrates the inner character of the shift: the splitting between the
cores is strong along the circumference of radius $r=1$, but is absent for $%
r=7$.

Systematically collected numerical results verify stability of the $S=1$
vortices with the global or inner shift between the cores of their
components. On the other hand, the well-known results for the
single-component GP/NLS equations \cite{Swartz,vortex} suggest that vortices
with $S\geq 2$ may be unstable against spontaneous splitting into sets of
unitary ones.

\section{Conclusion}

We have considered a system which models the copropagation of two circular
polarizations of light in the self-defocusing optical fiber with the linear
mixing of the polarizations, as well as the dynamics of the two-component
self-repulsive BEC with the Rabi coupling in 1D and 2D free-space settings
alike. A new effect which is demonstrated by the system is the SSB
(spontaneous symmetry breaking) in the bimodal CW state in the case when the
relative strength of the inter-component repulsion takes values $g>1$ (which
corresponds to the full immiscibility, in the absence of the linear mixing).
Exact analytical solutions for stable symmetry-breaking CW states are
obtained. Further, we have found the exact solution for stable 1D DS (dark
soliton) supported by the asymmetric (partly immiscible) CW background in
the case of $g=3$. For $g\neq 3$, generic symmetry-breaking DSs are found in
the numerical form. The full stability of these modes is established by
means of systematic simulations of their perturbed evolution. Furthermore, a
parameter region is identified in which the symmetric (fully miscible) CW
background supports stable DSs with the partial immiscibility of the two
components, in its core. The 2D system gives rise to stable vortex modes
with the shift (splitting) between the two components, which breaks the
vortex' isotropy. These modes include ones in which the global shift is
imposed by the asymmetric CW background, as well as modes with the symmetric
background, in which the local shift takes place between the vortical cores\
of the two components.

As an extension of the analysis, it may be interesting to consider
interactions between the symmetry-breaking DSs in 1D, and interactions
between the bimodal vortices with shifted components in 2D. It is also
relevant to extend the analysis to dissipative systems, modeled by coupled
complex Ginzburg-Landau equation \cite{Sigler}. In particular, experimental
observation of polarization SSB in dissipative solitons in a fiber laser was
very recently reported in Ref. \cite{new}.

\section*{Acknowledgment}

We appreciate valuable discussions with Eyal Buks. The work of B.A.M. was
supported, in part, by the Israel Science Foundation through grant No.
1695/22.

\end{document}